# Quantifying Jupiter's influence on the Earth's impact flux: Implications for planetary habitability.

Jonathan Horner[1], Barrie W. Jones[2]

[1] *Department of Astrophysics and Optics, School of Physics, University of New South Wales, Sydney 2052, Australia*
[2] *Astronomy Discipline, Department of Physics, Astronomy, and Space Science, The Open University, Milton Keynes, MK7 6AA, United Kingdom*

**Summary:** It has long been thought that the presence of a giant planet is a pre-requisite for the development of life on potentially habitable planets. Without Jupiter, it was argued, the Earth would have been subject to a punishing impact regime, which would have significantly retarded or outright prevented the development of life on our planet.

Although this idea is widely embraced, little research has previously been carried out to support it. Here, we present the results of several suites of dynamical integrations used to model the influence of Jupiter's mass and orbit on the impact rate that would be experienced by the Earth. We find that, far from being a simple shield, Jupiter's role in determining the terrestrial impact flux is significantly more complicated than previously thought. Far from being a simple friend, such giant planets are perhaps more likely to imperil the development of life on otherwise habitable planets.

## Introduction

Ever since it was first accepted that collisions with asteroids and comets could pose a significant threat to life on Earth, the idea that the impact rate of those bodies upon the Earth would be much higher, were it not for the protective influence of the giant planet Jupiter, has grown in the public and scientific conscience. Essentially, the hypothesis states that, without the particular mass and placement of Jupiter within our Solar System, the Earth would have been subject to a far more punishing impact regime throughout its evolution than has actually been the case, which would in turn have stymied or entirely prevented the evolution of life on Earth. The idea forms one of the cornerstones of the "Rare Earth" hypothesis, which argues that the Earth, as a habitable planet, is an incredibly rare or even unique place, and that therefore life should be incredibly scarce, or even non-existent, elsewhere in the Universe [1].

In recent years, with the ongoing success of the search for exoplanets, the question of planetary habitability has moved back to the forefront of astronomical research. In the next few years, the *KEPLER* satellite will detect the first truly Earth-like planets around other stars [2], and the search will be on for the first detection of life beyond our Solar System. But which of the many exo-Earths that are found should be prioritised in that search? The estimated impact flux upon the newly discovered exo-Earths will doubtless be a key metric in determining which planets to target in that search [3][4].

To date, it has simply been assumed that the presence of a giant, Jupiter-like, planet orbiting beyond any exo-Earth would be a requirement for life to develop and thrive [1][5]. However, little work has been carried out in the past to attempt to verify that hypothesis [6][7]. Here, we present the results of a series of detailed dynamical studies performed to attempt to address this lack of rigorous support for the "Jupiter – Friend" hypothesis. For a more detailed discussion of our work, we direct the interested reader to [8][9][10][11][12][13].

**Small bodies in the Solar system – three potentially threatening populations**

It is now well accepted that the Earth has been bombarded by asteroidal and cometary bodies throughout its history (e.g. [28][44][45][46]). When it was first acknowledged that impact craters on Earth were truly the scars resulting from collisions between the Earth and other Solar System bodies, the great majority of known objects on Earth-crossing orbits were long-period comets. For those objects, it was well known that a significant fraction are ejected from the Solar System entirely due to distant perturbations by Jupiter[1], and so it is easy to understand where the idea of "Jupiter – the Friend" originated. Simply put, were Jupiter less massive, fewer of the long-period comets would be removed from the Solar System, leaving a significantly greater number moving on orbits that could threaten the Earth (e.g. [6]).

However, in the decades since impact craters were first universally recognised as such, our understanding of the Solar System has undergone a dramatic change. Where the long-period comets were once considered by far the most significant source of potential impactors on the Earth, it is now accepted that they contribute only a small component of the total flux. At the current epoch, it is thought that there are three key populations of small bodies that pose a threat to the Earth: the near-Earth asteroids (e.g. [27][28][47]), the short-period comets (e.g. [17][19][25]), and the long-period comets ([14][48]). Each of these populations has different origins, and the role played by Jupiter in determining their evolution varies from case to case.

As discussed above, the long-period comets are currently thought to constitute the smallest contribution to the impact flux at Earth. These objects move on orbits that take thousands, or even a few million, years to complete, and their numbers are continually replenished by an ongoing injection of fresh comets from a vast reservoir known as the Oort cloud, containing as many as $10^{13}$ cometary nuclei held in cold storage, the great majority of which are smaller than 10km in diameter. The great bulk of these stored comets occupy a thick shell surrounding the Sun, between approximately $10^3$ and $10^5$ AU distant [14][15]. Members of the cloud are perturbed by a number of mechanisms, principle among them the effects of the galactic tide, with a contribution to the flux resulting from gravitational tweaks by passing stars. These perturbations result in the Oort cloud objects acquiring orbits that penetrate the inner Solar System, becoming dynamically new long-period comets [16].

The short-period comets represent a dynamically distinct population of cometary bodies that threaten the Earth. Rather than moving on orbits with periods of thousands or millions of years, such comets have orbital periods so short that they can be observed returning time and time again. The great bulk of such comets have orbital periods of around five or six years, with aphelia located in the vicinity of Jupiter's orbit, and as such are often referred to as the Jupiter family comets. The proximate parent population of the short period comets is well established as being the Centaurs (e.g. [16][17][18]). The origin of the Centaurs, however, is still the subject of much debate, with contributions to that population likely coming from the Edgeworth-Kuiper belt [19][20], Scattered Disk [21], inner Oort cloud [22], and even the Jovian and Neptunian Trojans [23][24][25]. Regardless of their origin, such comets are thought to contribute a significant fraction of the impact hazard at Earth, and might even represent the single largest component of that threat, dependent on the frequency with which giant Centaurs, such as Chiron, are captured to the inner Solar System [26].

---

[1] Indeed, in [14], Oort points to earlier work by Fayet, in 1929, stating *"A direct indication of the probable escape of a considerable fraction of the comets of very long period has been given by Fayet. Among 36 comets for which he has made approximate calculations of the orbits which they must have described after they passed out of the action of Jupiter he found 7 for which this orbit was hyperbolic."*

The final population of potentially hazardous objects, and that widely thought to contribute the bulk of the impact flux at Earth (fragmentation of giant comets aside) are the near-Earth asteroids. The great bulk of such objects are sourced to the inner Solar System from the main Asteroid belt [27], although some fraction of the population might well be de-volatilised short-period comets. Based on the number of near-Earth asteroids that have been discovered over the past few decades, some authors have suggested that the asteroidal contribution to the Earth's impact flux at the current epoch might be as high as three-quarters [27,28].

Given the distinct natures and origins of these three populations of threatening object, it is clearly important to consider the influence of Jupiter in determining the impact flux at Earth from each population in turn, particularly since the relative contribution of the three populations to the Earth's impact regime remains under debate. Such an approach also benefits the wider consideration of the role of giant planets in general in influencing habitability [3], since it is by no means certain that exoEarths would necessarily lie in systems with direct analogues of each of the three populations discussed. Understanding in a broader sense the influence of giant planets on impacts from an interior population of small bodies (in this case the near-Earth asteroids), and exterior disk of small bodies (evolving to become short-period comet analogues) and Oort cloud bodies will be vital in determining the impact regime for any newly discovered and potentially habitable exoEarths, particularly when the distribution of debris reservoirs within their host systems can to some extent be mapped by surveys such as the *Herschel* ([49]) Open Time Key Programmes DEBRIS [29] and DUNES [30], which generally detect cold disks analogous to the Edgeworth-Kuiper belt [50] (although those detected to date typically contain significantly more material than the reservoir in our Solar System). To detect the warmer dust present at astrocentric distances similar to that of our Asteroid belt, surveys at shorter wavelengths are necessary, and the detection of such dust typically results from observations using *WISE* and *Spitzer* data (e.g. [51][52]).

## The influence of Jupiter's mass

The simplest way to examine Jupiter's influence on the terrestrial impact flux is to consider the influence of Jupiter's mass. If Jupiter is purely a "friend", then one would expect that, were Jupiter more massive, then the Earth would experience fewer impacts, whilst if Jupiter were less massive, the impact rate would rise accordingly. In order to examine this hypothesis, we carried out detailed dynamical integrations using the *Hybrid* integrator within the *n*-body dynamics package *MERCURY* [31]. We considered Jupiter's influence on each of the three populations of potentially hazardous objects (the Asteroids [9], the short-period comets [10] and the long-period comets [11]) in turn.

The Near-Earth Asteroids

In order to examine the influence of Jupiter's mass on the impact rate at Earth resulting from the near-Earth asteroids, we ran simulations using *MERCURY* that followed the dynamical evolution of 100,000 mass-less test particles for a period of 10 Myr under the gravitational influence of the Earth, Mars, Jupiter, Saturn, Uranus and Neptune. The initial distribution of the test particles was set such that they were contained in "unperturbed" asteroid belts, an attempt to mimic the distribution of material that would be expected in the asteroid belt prior to its sculpting under the influence of Jupiter. For a detailed overview on how these belts were populated, we direct the interested reader to Appendix I of [9]. Because of the computationally intensive nature of the integrations performed, it was not feasible to study the dynamical evolution of a large enough sample of test particles that a significant number would impact the Earth, if its radius were kept at the canonical 6378km. In order to properly assess the impact flux, then, we had two options – either estimate the flux based upon the variation in the number of Earth-crossing asteroids as a function of time, or increase the effective cross-section of the Earth such that collisions would happen more frequently. We chose the latter option, and increased the radius of the Earth to 1 million kilometres. We then ran simulations for 12 distinct values of Jupiter's mass, ranging from 0.01 $M_J$ to 2.00 $M_J$, a process that took approximately 20 years of computation time, spread across the nodes of the Open University's IMPACT computing cluster. The impact rate as a function of Jupiter mass is shown in Figure 1.

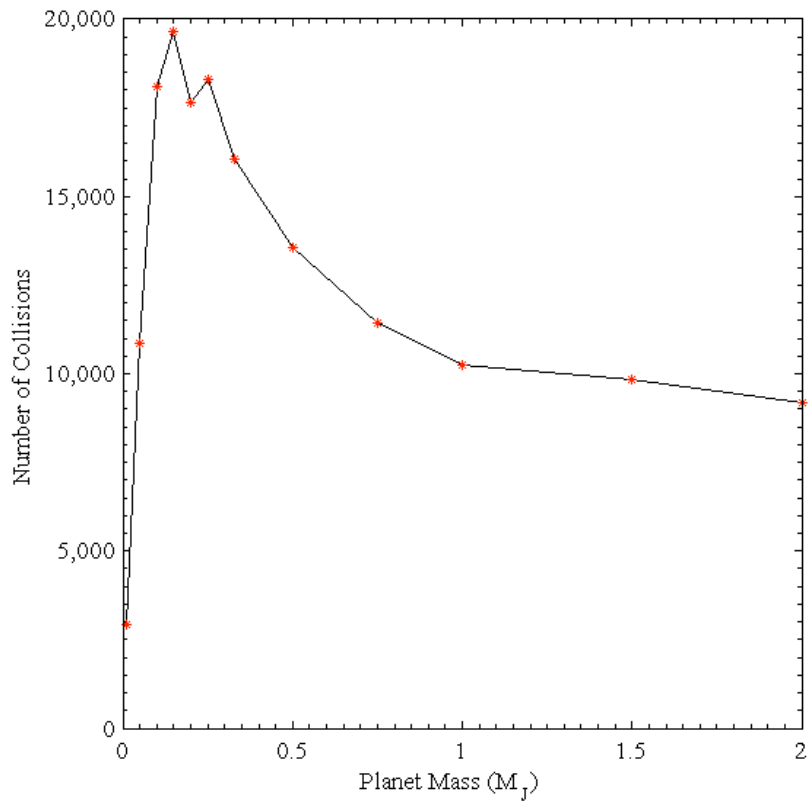

*Figure 1: The variation of the number of impacts from asteroidal bodies on the Earth over a ten million year period, as a function of Jupiter's mass. The initial population of test particles was 100,000, so in the worst-case scenario (between 0.1 and 0.25 $M_J$) almost 20% of the test particles collided with our inflated Earth, of radius 1 million kilometres.*

Two things are immediately clear from Figure 1. Firstly, the impact rate experienced by our inflated Earth at 1.00 $M_J$ is some 3.5 times greater than that when "Jupiter" has a mass of 0.01 $M_J$ – the giant planet is hardly a shield, in that case! In addition, the impact flux reaches a peak between 0.10 and 0.25 $M_J$, at which point the impact flux is almost twice that for the 1.00 $M_J$ case. So whilst our Jupiter is clearly not simply a "friend", it is clear that were Jupiter significantly less massive (say, closer to the mass of Saturn), then the impact flux at Earth from near-Earth asteroids would be elevated significantly over that we observe today. The reason for the enhancement for scenarios where Jupiter's mass is comparable to, or slightly smaller than, that of Saturn is the influence of the $v_6$ secular resonance. Within our own Solar

System, the destabilising influence of that resonance is well known as a source of fresh material from the inner reaches of the asteroid belt to the near-Earth asteroid population. As the mass of Jupiter is reduced, the location of the secular resonance moves outward, allowing it to destabilise a far greater fraction of the belt. The maximum destabilisation, and therefore the mass at which the most asteroidal material is flung in to the inner Solar System, occurs between 0.10 and 0.25 $M_J$.

The short-period comets

As for the near-Earth asteroids, we examined the influence of Jupiter's mass on the impact flux at Earth due to the short-period comets by carrying out detailed dynamical integrations that followed the evolution of just over $10^5$ test particles. Those particles were initially located on Centaur-like orbits that crossed the orbit of Neptune, but came no closer to the Sun than the orbit of Uranus. Rather than create a totally hypothetical population of test particles, we based our population on the orbits of 104 known Centaur and Scattered Disc objects, using each as a nominal orbit on which to base the orbits of 1029 independent test particles, each of which was created with orbital elements displaced somewhat from the orbit of the parent body[2]. As the Centaurs are the proximate parent population for the short-period comets, the chaotic diffusion of these test particles gave a continuous flux of fresh material to our test "Jupiters". Once again, we followed the evolution of the test particles for a period of $10^7$ years, and counted the impacts upon our inflated Earth. The results can be seen in Figure 2.

---

[2] It might initially seem that basing our population on just 104 parent objects might be somewhat risky (since it could introduce some biases in the resulting short period comet population). However, the clones of the 104 parent "seeds" were distributed very widely around the parent itself (far more so than would be standard if we were carrying out a study of that particular object). Furthermore, and as shown in [16], [17], and [18], objects placed on Centaur-like orbits rapidly diffuse dynamically across the entire outer Solar System, as a result of the perturbations of the four giant planets. As such, at the point of their injection into the inner Solar System, our test particles will be moving on greatly different orbits to those they initially occupy. Ideally, we would have had a situation where a greater number of parent objects had been discovered before we embarked upon our study, but given the wide dispersion of the clones used, together with their subsequent dynamical diffusion throughout the outer Solar System, we believe our test population to be as fair a representation of the parent population of short period comets as possible. For a detailed discussion, we direct the interested reader to [10].

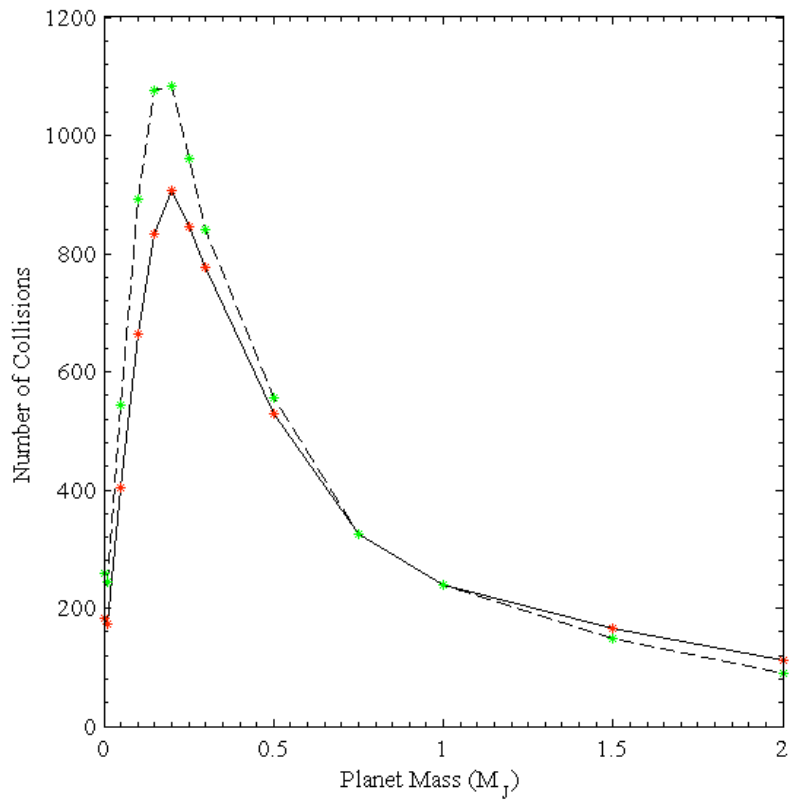

*Figure 2: The variation in the number of impacts experienced by an inflated Earth as a function of Jupiter's mass, for a population of just over 100,000 test particles based upon 104 Centaur and Scattered Disc objects. The points in red, joined by the solid line, show the number of collisions observed in our integrations, whilst those in green, connected with the dashed line, reproduce those results with the collision rate scaled according to the dynamical half-life of the Centaurs population in that particular integration.*

As can be seen from Figure 2, the impact flux at Earth is almost the same for a scenario in which Jupiter has little or no mass as that observed for the 1.00 $M_J$ scenario. In other words, the Earth experiences essentially the same impact flux from short-period comets in our Solar System as it would were Jupiter not present. However, the impact flux is once again markedly higher for scenarios between 0.10 $M_J$ and 0.25 $M_J$. In this case, the reasoning for this peak in the impact flux is rather straightforward. At high Jupiter masses, the planet is remarkably efficient both at flinging short-period comets onto Earth-crossing orbits, and at removing them from those orbits and ejecting them from the Solar System. At the lowest Jupiter masses, it can fulfil neither role with any efficiency. As a result, in both cases, the impact rate experienced at Earth is low. As the mass of Jupiter is increased from the lowest values, it first becomes sufficiently massive to inject fresh short-period comets to Earth-crossing orbits. As its mass increases further, it becomes ever more efficient at this task, whilst initially having great difficulty in removing those comets from the inner Solar System. As a result, the impact rate rises as a function of mass. Eventually, Jupiter becomes sufficiently massive that it can efficiently remove short-period comets from the inner Solar System. At this point, the impact rate begins to fall once again, as Jupiter becomes ever more efficient at disposing of potentially hazardous comets.

The data plotted in green in Figure 2 simply adjusts the total number of impacts obtained in our integrations for the size of Centaur population that would be expected in those runs. Given that the flux into the Centaur region is, at least to first order, independent of Jupiter's mass, whilst the rate at which objects are removed from the Centaur population is heavily dependent on Jupiter's mass, it is clear that the total population of Centaurs will be larger the less massive Jupiter is. In plotting the green points, then, we use the bulk data from our simulations to calculate a dynamical half-life for the Centaur population in each case, and use that dynamical half-life to rescale the results to take account of this effect. As can be seen, this has only a small effect on our results. In fact, the main consequence of this correction is that the impact rate experienced by Earth for scenarios with the least massive "Jupiter"s is almost exactly the same as that for the 1.00 $M_J$ scenario (rather than being slightly smaller).

The long-period comets

To investigate the influence of Jupiter on the impact rate from the long-period comets in the manner described above would require unfeasible computational power. Since the orbital periods of dynamically new Oort cloud comets (those passing through the inner Solar System for the first time) are typically measured in hundreds of thousands, or even millions of years, the number of times any given comet will pass through the inner Solar System in a given ten, or even hundred, million year window will be relatively small. This in turn means that the likelihood of any given comet colliding with the Earth, even if the Earth is inflated, over that timescale would be vanishingly small. As such, we needed to take a different tack to examine Jupiter's role in shielding the Earth from the long period comets. For those comets, we chose to examine the fraction surviving as Solar System bodies as a function of Jupiter's mass as an initial proxy for the impact rate as a function of Jovian mass. As a further additional test, we took account of the orbital periods of the surviving comets as a function of time, and used that, in conjunction with the number surviving, to estimate the impact flux upon Earth as a function of time in each simulated scenario. For simplicity, we again followed the orbits of 100,000 test particles, spread on initial orbits appropriate for dynamically new long period comets. In this case, we helped speed the integrations along by considering only the gravitational influence of Jupiter and Saturn, and neglecting the effects of the other planets. While this represents a further simplification of the planetary system over our previous runs (in which the effects of Uranus, Neptune and the Earth were also included), it is clearly not an unreasonable approximation. We further simplified by considering a smaller sample of Jupiter masses – 0.00, 0.25, 0.50, 1.00 and 2.00 $M_J$. We followed the evolution of the test particles for a period of 100 Myr, and kept track of the number that remained on bound orbits through the runs, together with their orbital periods. For a detailed discussion of our methods and results, we direct the interested reader to [11].

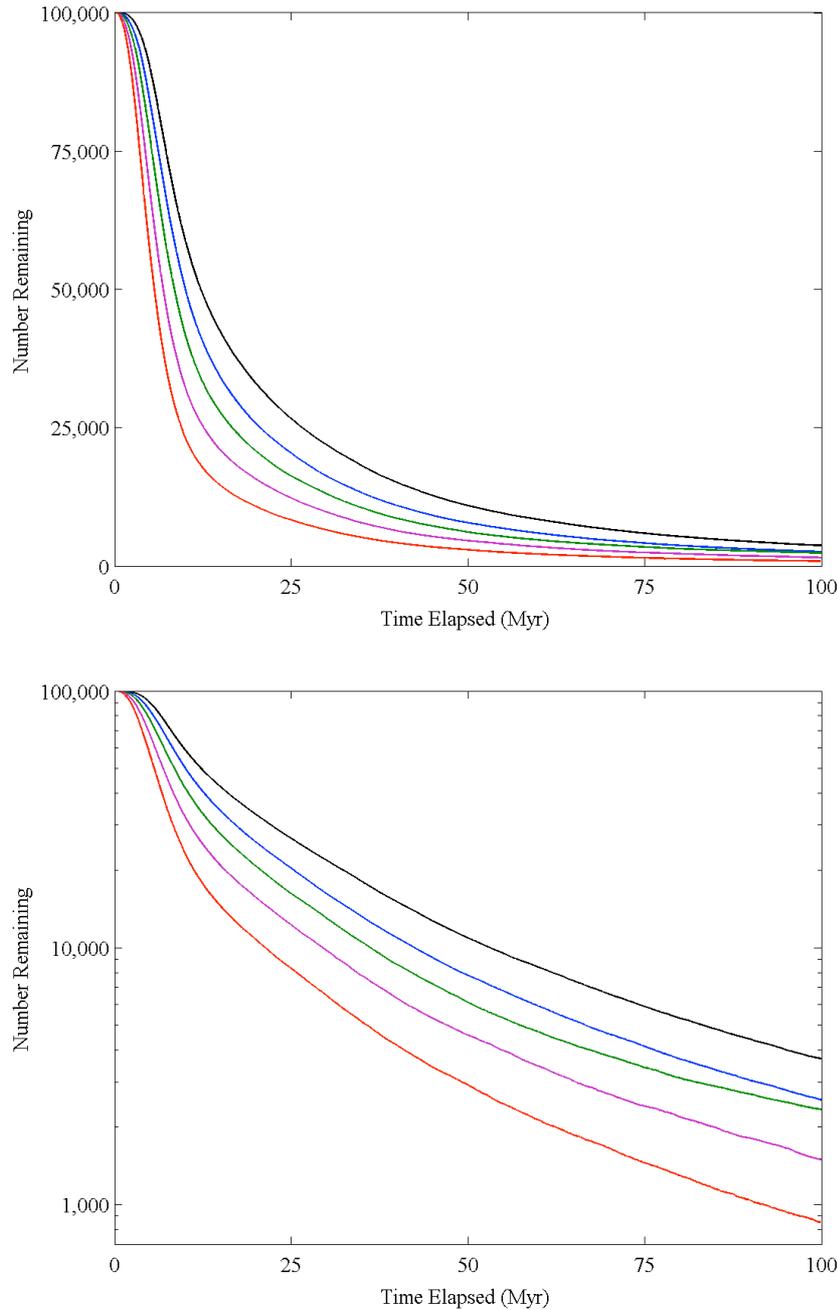

*Figure 3: The number of long-period comets remaining gravitationally bound to the Solar System as a function of time, for the five scenarios tested in this work. The results for the different "Jupiter" masses are plotted in different colours, as follows: 0.00 $M_J$, 0.25 $M_J$, 0.50 $M_J$, 1.00 $M_J$, and 2.00 $M_J$.*

As can be seen from Figure 3 and Table 1, the number of long period comets that remain gravitationally bound to the Solar System falls rapidly as a function of time, even when the mass of Jupiter is set to 0 (i.e. the only perturbing planet in the Solar System is Saturn). However, it is also clear that the number of surviving comets falls far more rapidly for scenarios in which Jupiter is more massive. This, in turn, means that fewer comets remain on Earth-crossing (and therefore Earth-threatening) orbits, and so one would expect the impact rate on the Earth to be significantly lower. In other words, for the long period comets, it seems that Jupiter truly is a "friend". This is no real surprise, since we believe that the idea that Jupiter acts as a shield to the Earth dates back to a time when such comets were considered by far the greatest impact threat to our planet. The role of Jupiter in ensuring that so many such comets are ejected from the system after only a few orbits (often on their first pass through the Solar System) is well accepted. However, we remind the reader that, at the current epoch, such comets are thought to only contribute a minor fraction of the impact flux experienced by the Earth. We note in passing that, even when the fact that Jupiter can also act to shorten the orbital period of long period comets (and therefore ensure that a given comet encounters the Earth more frequently than it otherwise would) is taken into account, we still find that the lowest impact rates on Earth from this population occur for the highest Jupiter masses.

| Mass ($M_J$) | Initial | 1 Myr | 10 Myr | 100 Myr |
|---|---|---|---|---|
| 0.00 | 100000 | 99982 | 58949 | 3689 |
| 0.25 | 100000 | 99861 | 50138 | 2551 |
| 0.50 | 100000 | 99681 | 41835 | 2337 |
| 1.00 | 100000 | 99314 | 32334 | 1495 |
| 2.00 | 100000 | 98659 | 23253 | 852 |

*Table 1: The number of test particles remaining gravitationally bound to the Solar System as a function of time and of Jupiter mass. Each simulation started with an identical population of 100,000 test particles, and the only difference between them was the mass of "Jupiter".*

## The influence of Jupiter's orbit

Whilst it is clear that Jupiter's mass plays a significant role in determining its influence on the impact flux at Earth, it is clearly interesting to also consider the influence of Jupiter's orbit. To follow our earlier work investigating the influence of Jupiter's mass, we therefore investigated the influence of the eccentricity and inclination of Jupiter's orbit on the impact flux at Earth from the near-Earth Asteroids and the Centaurs.

In our own Solar System, Jupiter moves on an orbit with relatively low orbital eccentricity, and very small inclination. As such, it clearly makes sense to ask what the impact regime at Earth would be if the planet's orbit was significantly more eccentric or more highly inclined than that we observe. This question is particularly relevant given the increasing catalogue of planetary systems that involve at least one massive planet on an orbit significantly more inclined or eccentric than those seen in our own Solar System (e.g. [32][33][34]).

In order to test the influence of Jupiter's orbit on the impact rate at Earth, we repeated our earlier work in its entirety with several new suites of orbital integrations. In every case, the initial conditions used were identical to those in the earlier work (so the same number of test particles, placed on the same orbits, with the same time steps and same massive bodies present in the integration). The only thing changed in each case was the orbit of Jupiter. We first performed some simple trials to find the limits beyond which modifying Jupiter's orbit would drive the orbits of the Solar System's planets into an unstable configuration. Clearly, if the planetary system itself is unstable, then the question of a giant planet's influence on the impact rate on an otherwise habitable world becomes somewhat moot! As a result, we found that increasing the orbital eccentricity of Jupiter significantly beyond 0.1 caused major instability on timescales of a few million years (or even less). We therefore chose to model four new scenarios, two each in orbital eccentricity and inclination space.

To test the influence of Jupiter's orbital eccentricity, we considered scenarios in which the planet's initial orbital eccentricity was somewhat higher, and somewhat lower, than that we observe in our Solar System. In those runs, as detailed above, the only variable changed from our earlier work was the eccentricity of Jupiter's orbit. For the high eccentricity case, a value of 0.1 was chosen, whilst a value of 0.01 was used for the low eccentricity case. These neatly bracket the actual value of Jupiter's orbital eccentricity used in our earlier work, 0.048775.

When testing the influence of Jupiter's orbital inclination, we were aware that, within our own Solar System, Jupiter's orbit lies very close to the invariable plane of the system (with a current value of just 1.31°). We therefore opted to test moderate (5°) and high (25°) orbital inclinations

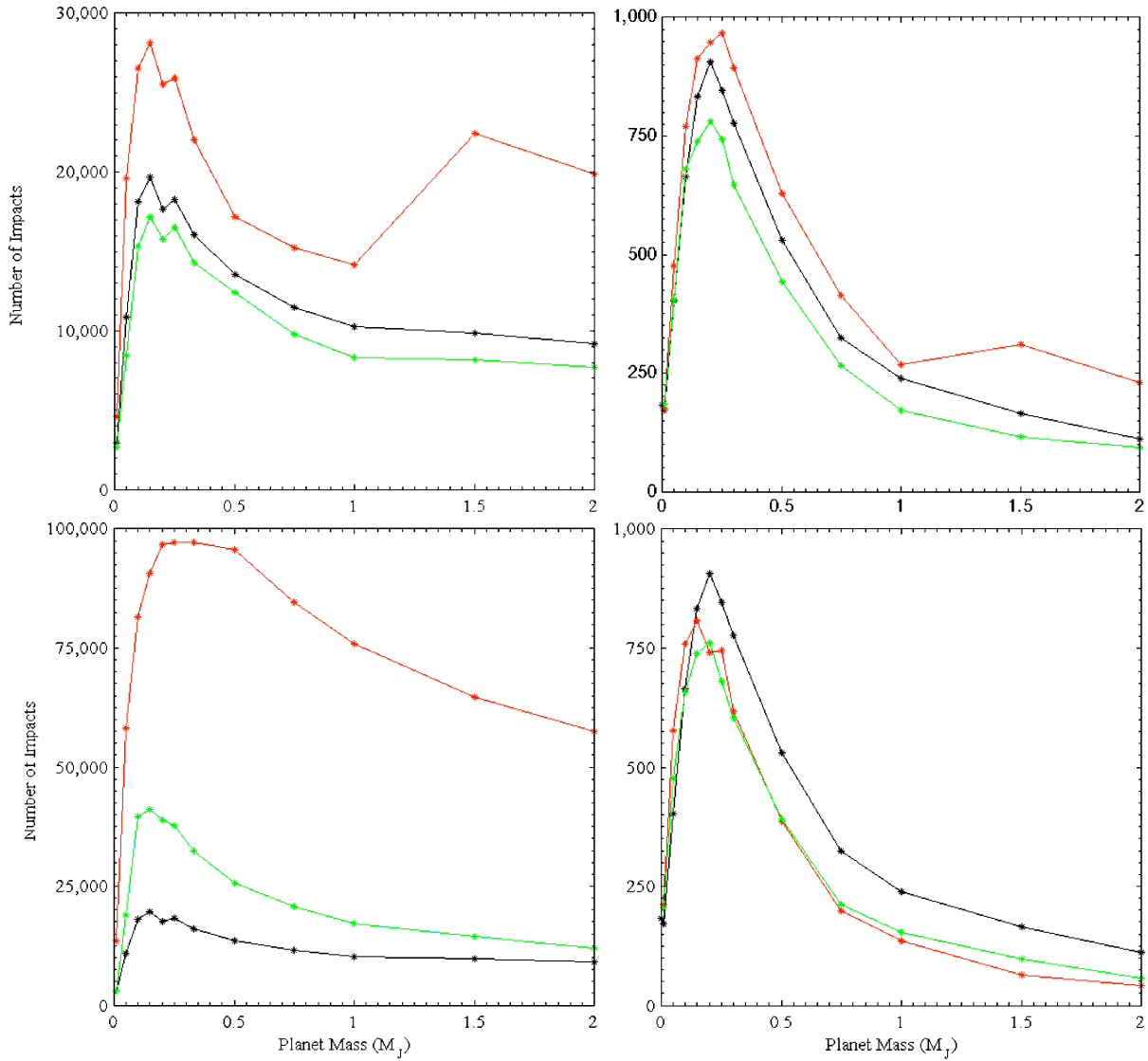

*Figure 4: The influence of Jupiter's orbital eccentricity (upper plots) and orbital inclination (lower plots) on the impact flux at Earth, as a function of Jovian mass. In the case of the plots showing the effect of orbital eccentricity (top), the line in red shows the results for an orbital eccentricity of 0.1, and those in green the results for an eccentricity of 0.01. For the plots showing the influence of orbital inclination, the green line details the results for a moderate inclination (5°), whilst those in red show the results for a high inclination (25°). The left hand plots show the results for the near-Earth asteroids, and those on the right show the equivalent results for the short-period comets. We note that, in the high inclination case for the asteroidal population (lower left), over 90% of the test particles collided with the inflated Earth at some point in the simulation for masses between 0.15 and 0.50 $M_J$.*

The results of our integrations can be seen in Figure 4. In that Figure, the results of our simulations involving the near-Earth asteroids are shown in the left-hand column, and those for the short-period comets are shown to the right. The upper plots show the results for our tests of orbital eccentricity, and the lower the results of orbital inclination.

For the runs that examined the influence of Jovian eccentricity, it is clear that the more eccentric the giant planet's orbit, the greater the impact flux that would be observed from the Earth. In the case of the results for the near-Earth asteroids, this effect is far more pronounced than in the case for impacts from short-period comets. In addition, a secondary peak in the impact flux can be seen in the asteroidal results at 1.50 $M_J$. Whilst it is clear that the eccentricity of Jupiter's orbit clearly plays a role in determining the impact flux at Earth, with more eccentric orbits yielding more impacts, that effect seems to be secondary to the role played by Jupiter's mass.

For the integrations that studied the influence of Jupiter's orbital inclination, the situation is somewhat different. For the short-period comets, it seems that Jovian orbital inclination plays only a minor role in determining the impact flux at Earth, although there is evidence that that flux is slightly higher for the scenario in which Jupiter lies closest to the system's invariable plane. For the near-Earth asteroid impact flux, however, the results are dramatic. A moderate increase in Jupiter's orbital inclination is enough to significantly increase the impact flux at Earth. However, in the scenario in which Jupiter's inclination was initially 25°, the asteroid belt was severely disrupted across almost the entire range of masses tested, with over 50% of the test particles colliding with the inflated Earth for all scenarios except that with the least massive Jupiter (0.01 $M_J$). In particular, for those runs between 0.15 and 0.50 $M_J$, the fraction of test particles colliding with the inflated Earth was well over 90%. Whilst this might initially suggest that exoEarths in planetary systems featuring inclined giant planets would be bad places to search for life, we contend that the incredibly high impact rates featured in those runs are a sign that the presence of such a highly inclined Jupiter-like planet would entirely remove the asteroid belt on an astronomically short timescale, with the end result that the long-term impact flux from asteroidal bodies would be low, simply because so few would remain!

## Conclusion

We have performed a series of detailed dynamical simulations aimed at answering the question "Jupiter – Friend or Foe?" Although it is widely held that Jupiter has acted to shield the Earth from impacts since its formation, and has therefore played a pivotal role in determining the habitability of our planet, little work had previously been done to investigate whether that is actually the case.

When it comes to the impact hazard posed by the long-period comets, which are sourced from the Oort cloud, the situation is fairly clear-cut. There, we find that Jupiter does offer a significant amount of shielding as a result of the efficiency with which it ejects such comets from the Solar System, never to return. However, it is generally accepted that such long-period comets constitute only a minor contribution to the total impact flux at Earth (perhaps of order 5%), and so this result does not, necessarily, infer that the old idea of "Jupiter – the Friend" is justified.

When we consider the influence of Jupiter in protecting (or imperilling) the Earth from impacts involving near-Earth asteroids and short-period comets, we find that, rather than simply being a shield or a threat, Jupiter's role is actually somewhat more complicated. In each case, the impact rate from such objects is markedly lower for planetary systems that include a massive Jupiter (such as our own) than for those that have a Saturn-mass (or slightly smaller) planet at the same location. However, for masses lower than ~0.15 times that of our Jupiter, the impact flux experienced by an Earth-like planet falls dramatically in both cases, such that the impact rate were no Jupiter present (or only a very-low mass planet occupied

Jupiter's orbit) would actually be lower than that for the scenarios involving our Jupiter. As such, it seems that Jupiter can easily be at least as much, if not more, of a foe than it is a friend.

We also considered the influence of the orbit of Jupiter on the impact rate at the Earth. We found that, were Jupiter's orbit somewhat more eccentric than that we observe, the impact flux at Earth would be somewhat greater, albeit not punishingly so. As such, we consider that the orbital eccentricity of a giant planet such as Jupiter is of secondary importance, compared to its mass, in determining the impact flux within its host system. For objects sourced from beyond the orbit of the giant planet, we found that the inclination of that planet's orbit has little or no effect on the impact regime in the inner reaches of the system. However, we found that increasing the planet's orbital inclination had a dramatic effect on the dynamical stability of the asteroid belt, leading to greater than 50% depletion of that belt on timescales of just ten million years for all but the least massive "Jupiter" tested (0.01 $M_J$). Whilst such destabilisation might initially cause one to consider such systems very unlikely to host a habitable planet, we contend that the level of destabilisation observed is sufficient that few objects will remain within that reservoir to pose a threat on astronomically significant timescales, and as such, those systems would simply not contain an asteroid belt to threaten any exoEarths of interest.

The simple notion that giant planets are required to ensure a sufficiently benign impact regime for potentially habitable worlds to be truly habitable is clearly therefore not valid. There remains, also, the question of what level of impact flux is more suitable for the development of life, and of advanced life. On one hand, the Earth, and other potentially habitable planets, will typically form interior to the "ice line", the location in a protoplanetary disk at which water can first be found as a solid. Interior to the ice line, the only water present in a solid form would be that trapped in hydrated silicates, and as such, one would expect such planets to form "dry", particularly if they only accrete from their local environment. The origin of the Earth's oceans, then, is still heavily debated. A detailed review of the debate surrounding the hydration of potentially habitable worlds is beyond the scope of this work (and we direct the interested reader to p.285 of [3] for more information), but it is certainly interesting to consider the influence of giant planet's on the impact flux at Earth-like planets in this context. If we accept that the bulk of the Earth's water was delivered from beyond the ice-line (either from wet asteroids, or from comets – e.g. [35]), then it follows that the delivery of that icy material must have primarily been driven by Jupiter. Does this mean that, if Jupiter were not present, or were of particularly low mass, then the Earth would not have received sufficient hydration to currently be habitable? On the other hand, if Jupiter had a mass similar to that of Saturn (and therefore the impact rate of trans-Jovian material were higher), would that result in a super-saturated "ocean Earth", which might not prove so clement for life? Unfortunately, whilst such discussion is clearly interesting, it is almost certain that the hydration of the Earth occurred during the migration of the outer planets, which led to significant destabilisation and redistribution of the Solar System's small body populations (e.g. [36][37][38][39][40]). As such, it is unclear whether our results (which are based on a stable, post-migration planetary system) can throw any light on this particular aspect of the problem.

Once Earth-like planets have been hydrated, the role of impacts will clearly shift from having import in the delivery of volatiles to otherwise dry worlds to directly affecting the course of the development of life. Since the development of life on our planet, a significant number of "mass extinctions" have occurred, in which the great majority of organisms have been extinguished. Although many of these are currently believed to been caused by factors other than impacts, at least a few are thought to have been at least partially the result of collisions between the Earth and small bodies (e.g. [41][42][43]). At first glance, it seems reasonable to assume that the most promising conditions for life to develop, once a host planet has received sufficient hydration, would be those featuring the lowest impact rate (i.e. those with the least massive giant planets, no giant planets at all, or very massive giant planets). However, it could equally be argued that at least *some* impact flux is necessary in order to trigger occasional mass extinctions – without the mass extinction that wiped out the dinosaurs, for example, it is debatable whether we would currently be here, debating the importance of such extinctions! Perhaps, then, we face a "Goldilocks" style situation, where the optimal impact flux is neither too high nor too low.

It is certainly clear that the role of giant planets in influencing the impact flux (and therefore the habitability) of Earth-like planets is significantly more complicated than was once thought. When the first truly Earth-like planets are detected around Sun-like stars, and the search for life upon them becomes feasible, it is clearly therefore imperative that the potential impact regime of each is addressed in some detail, in order to determine which of those planets represents the best targets for that search, rather than simply looking for those with the good (or bad!) fortune to move in planetary systems that involve distant giant planets.

## Acknowledgements

JH gratefully acknowledges the financial support of the Australian government through ARC Grant DP0774000. The authors also wish to think the anonymous referees, whose suggestions helped to improve the flow and background content of our article.